\documentclass[useAMS,usenatbib]{mn2e}

\usepackage{graphicx,epsfig,subfigure}
\usepackage{url}

\usepackage{amsmath}
\usepackage{amssymb}

\usepackage{threeparttable}
\usepackage{float}
\usepackage{morefloats}

\usepackage{ifsym}

\usepackage{xcolor}

\title[The dynamical evolution of multiple systems of trapezium type]
   {The dynamical evolution of multiple systems of trapezium type}

    \author[Allen et al.]
  {Christine Allen,$^1$\thanks{E-mail: chris@astro.unam.mx}
 Alex Ruelas-Mayorga,$^1$ Leonardo J. S\'anchez,$^1$ Rafael Costero,$^1$
  \\
    $^1$Instituto de Astronom\'{\i}a, Universidad Nacional Aut\'onoma
     de M\'exico, Cd. Universitaria, M\'exico, D.F. 04510, M\'exico\\}

\begin{document}

\date{Accepted 2018. Received 2018 June; in original form 2018 June}

\pagerange{\pageref{firstpage}--\pageref{lastpage}}
\pubyear{2018}

\maketitle

\label{firstpage}


\begin{abstract}

We have selected archival observational data for several O and B trapezia in the Milky Way. For each of the main components of the trapezia we obtained  transverse velocities from the historical separation data. With this information, and with the stellar masses of the main components, we studied the dynamical evolution of ensembles of multiple systems mimicking each one of the trapezia. For this purpose we conducted numerical $N-$body integrations using the best available values for the masses, the observed positions and {\bf transverse} velocities, randomly generated radial velocities, and random line-of-sight ($z$) positions for all components. Random perturbations  were assigned to the observed quantities, compatible with the observational errors.  A large fraction of the simulated systems ({\bf between 70 and 90 percent}) turned out to be unbound. The properties of the evolving systems are studied at different values of the evolution time. We find that the dynamical lifetimes of both the bound and unbound seems to be quite short, of less than 10~000 yr for the unbound systems, and of 10~000 to 20~000 yr for the bound systems. The end result of the simulations is usually a  binary {\bf with semiaxes of a few hundred AU}  sometimes a triple of hierarchical on non-hierarchical type. Non-hierarchical triples formed during the integrations (which are dynamically unstable) were found to have much longer lifetimes, of 250~000 to 500~000 yr. The frequency distributions of the major semi-axes and eccentricities of the resulting binaries are discussed and compared with observational properties of binary systems from the literature.

\end{abstract}




\section{Introduction} 
\label{sec:intro}

A recent study of the dynamical evolution of the Orion Trapezium (OT) showed that it is probably very unstable and not likely to survive for more than 10~000 years, unless the combined mass of component C was about 65 $M_\odot$ \citep{Allen2017}.  With this assumption, the lifetime of the system turned out to be between 10~000 and 50~000 years.  A previous study of the minicluster $\theta^1$~Ori B  \citep{Allen2015} also showed a short dynamical age for this system, of about 30~000 years, compatible with that of the entire Trapezium only if we assume the larger mass for component C. These results motivated us to look in the literature for additional data for multiple systems of trapezium type in order to study whether their dynamical evolution would be similar to that of the prototype, the Orion trapezium.

To our knowledge, studies on the dynamical stability of trapezium-type systems still remain scarce in the literature. Examples include \cite{Allen1974a, Pflamm2006}, and \cite{Allen2017}. \citet{Allen1974a} studied a sample of 30 massive systems, simulating their dynamical evolution by N-body integrations.  These systems had  trapezium-like initial positions and virialised velocities.  Their main conclusion was that after 30 crossing times (corresponding to about 1 million years) only  30\% of the systems survived as trapezia. In another study, \cite{Pflamm2006} found that the probability of survival of a trapezium after 5 000 years was 14\% and that after $10^6$ years only about 4\% of the trapezia survived as such. The results of these two studies, though, are not directly comparable, because of different assumptions, parameters, and definitions in both.

{ \bf We follow the definition of \citet{Ambartsumian1954} for a trapezium.  Let a multiple star system (of 3 or more stars) have components a, b, c, … and let ab, ac, bc, .. be the distances between the components. If three or more distances are of the same order of magnitude, then the multiple system is of trapezium type;  if no three distances are of the same order of magnitude, then the system is of ordinary type. (For the sake of clarity, we shall call these systems hierarchical systems). Two distances are of the same order of magnitude in this context if their ratio is greater than 1/3 but less than 3. 
Normally, only the separations projected on the plane of the sky are available.  Using these, the sample of observed trapezia will contain, in addition to true trapezia also optical trapezia and pseudotrapezia. An optical trapezium is an apparent multiple system whose components are not physically connected, but appear close together due to  projection. A pseudo-trapezium is a multiple system of hierarchical type that appears to be a trapezium due to projection. Based on data then available, Ambartsumian showed that the number of pseudotrapezia is small only among multiple systems of spectral types O and B. As we shall see later, \citet{AbtCor2000} were able to refine and update this analysis.}  
 
A comprehensive search in the literature provided us with observational data on 29 trapezium-type systems. However, not all of them proved suitable for numerical modelling. We discuss in Section \ref{sec:selectcrit} the criteria we used for the selection of the trapezium systems suitable for this dynamical study. Section \ref{subsec:transvel} presents the data for the transverse velocities and their uncertainties. Section \ref{subsec:masses} is devoted to an examination of the available material on the masses of the individual trapezium members. Section \ref{subsec:init-con} describes the initial conditions taken and the numerical integrations performed to model the dynamical evolution of ensembles of trapezium-like systems. The general outcome of the modelling of the individual trapezia of spectral type O is presented and discussed in Section \ref{subsec:otrap}, that of the B-type trapezia in Section \ref{subsec:btrap}. Section \ref{subsec:unbound} discusses the unbound systems. Section \ref{subsec:lifetimes} is devoted to an analysis of the lifetimes of the modelled systems. Section \ref{subsec:ejectedstars} presents the velocity distribution of the ejected stars, and Section \ref {subsec:binariesformed} deals with the properties of the binaries formed during the integrations. Finally, in Section \ref{sec:conclusions} we summarise our results and conclusions.


\section{Trapezia Selection Criteria}
\label{sec:selectcrit}

The trapezia used for this investigation were selected from several sources \citep{Allen1974b, Allen2004, Abt1986, AbtCor2000}.  Our first selection criterion was that data on the proper motions of the primaries were available. This enabled us to ascertain whether, and which, of the  secondaries were likely to be physical members of the systems. In this way we were able to eliminate systems with optical components corresponding to stars that appear close the trapezia, but are located at different distances and hence show discrepant proper motions.  
Careful studies were conducted by \citet{Abt1986} and \citet{AbtCor2000}, by means of CCD photometry, spectroscopy and astrometry. They observed the components of candidate trapezium systems from an unpublished catalogue by Allen. They found, among other things, that the magnitudes listed in the IDS were systematically too bright. They measured angular separations and position angles between components, and determined accurate stellar magnitudes;  they also constructed colour-magnitude and colour-colour diagrams which, along with spectroscopy, allowed them to estimate distances to the individual components of the systems, and thus to discard foreground and background stars as non-members. Their final list included 28 candidate trapezium systems.  
A previous study by \citet{Ambartsumian1954} showed that about 9\% of the hierarchical multiple systems should appear as trapezia in projection. These pseudo-trapezia, like other hierarchical systems, have a broad distribution of primary spectral types. \citet{Ambartsumian1954} concluded that the systems most likely to be true trapezia were those with primaries of spectral types O and B. \citet{Abt1986, AbtCor2000} were able to refine this criterion, and concluded that, among their list, the systems most likely to be true trapezia were those with primaries of spectral types B3 and earlier, plus two later-type evolved systems that, while on the main sequence, had spectral types B3 and B1. However, these two systems do not appear in their Table 3, so no further information on them is available. Only 12 of the remaining systems in \cite{AbtCor2000} are of type B3 and earlier.  We adopted these 12 systems as our primary material, and tried to increase the sample with 3 trapezia considered to be physical by \citet{Allen1974b}. However, two of these turned out to have spectral types later than B3, so they were discarded.  This process left us with a list of 13 systems which, according to the criteria listed, are likely to be  true trapezia,
Next, we requested the historical data on separations and position angles for these 13 trapezia from the WDS catalogue \citep{Mason2001}, which were kindly supplied by B. Mason, and from which we estimated the angular separation velocities. With the best available distances for these systems (mostly by  \citet{Abt1986, AbtCor2000} ) we were able to calculate the separation velocities in kilometres per second. Some of these turned out to be very large, larger than about 50 km/s, thus exceeding by much the plausible escape velocities.  We took this as an indication that the corresponding component was most likely an optical companion.  In this way we excluded the following systems:  ADS 15834, ADS 15387, ADS 13368.  This left us with a final sample of 10 trapezia,  three with primaries of spectral type O, seven with B3 or earlier primaries.

{\bf The recent publication of the Second Gaia Data release contains a plethora of data, but its usefulness for the analysis of the kinematical and dynamical behaviour of trapezium systems is still to be ascertained. Potential problems include the presence of close companions (which are treated by Gaia as single stars) and many possible undetected binaries. Furthermore, no radial velocities for the trapezium components are yet available. Since the primaries of or systems are all of early spectral type it is uncertain whether they will become available in the near future. At the present time, we did not find Gaia data relevant for our study.}

The list of the systems we will study is presented in Table \ref{tab:trapeziatable}. The first four columns of this table give different catalogue identifiers for the trapezium in question; the last column lists the spectral type of the primary star.
 

\begin{table*}
\caption{List of selected trapezia with ID cross references and spectral type of the primary star.}
\begin{center}

\begin{tabular}{ccccc}

\hline

ADS		&	WDS			&	IDS			&	BDS		& 	Primary type	\\      
\hline        
		&				&				&			&			\\
719		&	00528+5638	&	00470+5605	&	455		&	O5V		\\
1869	&	02291+6207	&	02215+6140	&	....	&   B2Vn	\\
2843	&	03541+3153	&	03478+3135	&	1921	&	B1I		\\
4728	&	06085+1358	&	06028+1359	&	3171	&	B1V		\\

10049	&	16256-2327	&	16196-2313	&	7613	&	B2V		\\
		&				&				&			&			\\
11169	&	18138-2104	&	18078-2105	&	8413	&	B9Ia	\\

13292	&	20024+3519	&	19586+3502	&	9944	&	B1Vn	\\

13374	&	20060+3547	&	20022+3530	&	9916	&	WN5+O9I	\\

15184	&	21390+5729	&	21358+5702	&	11160	&	O7V		\\

16795	&	23300+5833	&	23254+5800	&	12405	&	B3V	\\
 		&               &               &			&		\\

\hline

\end{tabular}
\label{tab:trapeziatable}
\end{center}
\end{table*}


\begin{table*}
\caption{Characteristics of selected trapezia. The first group contains the O trapezia, the second the B trapezia.}
\begin{center}

\begin{tabular}{ccccccccr}
\hline
ADS	&	Distance     	&	Components	&	Position 	&	Relative Separation      	&	Relative Separation    	&	 $\Delta$ Relative Separation 	&	Separation Velocity     	 \\
    &    (pc)           &               & angle ($^{\circ}$) & (arcsec)         &      (AU)         &    ($\arcsec$/yr)     &    (km/s)           \\
\hline															
719	&	1500	&	AB	&	83	&	1.5	&	2283	&	0.0003	&	1.82 $\pm$ 3.6    	 \\
	&		&	AC	&	135	&	3.9	&	5847	&	-0.0007	&	     -4.66 $\pm$ 3.6    	 \\
	&		&	AD	&	194	&	8.9	&	13350	&	-1.02E-07	&	     -7.27e-04 $\pm$ 3.6	 \\
	&		&	AE	&	334	&	16.8	&	25140	&	0.0022	&	15.77 $\pm$ 3.6   	 \\
                	&		&		&		&		&		&		&		 \\
13374	&	1574	&	AB	&	65	&	6.8	&	10656	&	0.0023	&	17.40 $\pm$ 3.8   	 \\
	&		&	AC	&	32	&	11.8	&	18573	&	0.033	&	246.46 $\pm$ 3.8  	 \\
	&		&	AD	&	300	&	11.3	&	17796	&	0.0023	&	17.48 $\pm$ 3.8   	 \\
	&		&	AE	&	109	&	28.7	&	45174	&	-0.0021	&	-15.47 $\pm$ 3.8  	 \\
                	&		&	AF	&	28	&	36.0	&	56724	&	-0.0007	&	-5.12 $\pm$ 3.8   	 \\
                	&		&		&		&		&		&		&		 \\
15184	&	900	&	AB	&	320	&	1.7	&	1566	&	0.0014	&	6.10 $\pm$ 2.2    	 \\
	&		&	AC	&	120	&	11.9	&	10722	&	0.0011	&	4.74 $\pm$ 2.2    	 \\
                	&		&	AD	&	339	&	20.0	&	17996	&	0.002	&	8.69 $\pm$ 2.2    	 \\
	&		&		&		&		&		&		&		 \\
\hline															
                	&		&		&		&		&		&		&		 \\
1869	&	1260	&	AB	&	145	&	4.0	&	5017	&	5.67E-03	&	33.86 $\pm$ 3.0   	 \\
	&		&	AC	&	43	&	4.7	&	5893	&	0.0073	&	43.82 $\pm$ 3.0   	 \\
      	&		&		&		&		&		&		&		 \\
2843	&	240	&	AB	&	205	&	13.1	&	3142	&	0.0021	&	2.37 $\pm$ 0.6    	 \\
	&		&	AC	&	286	&	33.3	&	7992	&	0.0062	&	7.06 $\pm$ 0.6    	 \\
	&		&	AD	&	195	&	98.6	&	23664	&	0.0624	&	71.00 $\pm$ 0.6   	 \\
	&		&	AE	&	186	&	120.0	&	28800	&	-0.0064	&	-7.28 $\pm$ 0.6   	 \\
	&		&		&		&		&		&		&		 \\
4728	&	794	&	AB	&	111	&	2.4	&	1889	&	0.0086	&	3.24 $\pm$ 1.9    	 \\
	&		&	AD	&	122	&	28.0	&	22230	&	-0.0024	&	-9.00 $\pm$ 1.9   	 \\
	&		&	AE	&	185	&	44.1	&	35022	&	0.0004	&	1.54 $\pm$ 1.9    	 \\
	&		&		&		&		&		&		&		 \\
10049	&	158	&	AB	&	334	&	3.0	&	481	&	-3.81E-03	&	-2.86 $\pm$ 0.4   	 \\
	&		&	AC	&	0	&	149.2	&	23578	&	-0.0137	&	-10.24 $\pm$ 0.4  	 \\
	&		&	AD	&	252	&	156.4	&	24706	&	-0.001	&	-0.74 $\pm$ 0.4   	 \\
                	&		&		&		&		&		&		&		 \\
11169	&	1150	&	AB	&	258	&	16.9	&	19401	&	1.97E-04	&	1.07 $\pm$ 2.8    	 \\
	&		&	AC	&	119	&	25.5	&	29325	&	5.01E-03	&	27.33 $\pm$ 2.8   	 \\
	&		&	AD	&	312	&	48.2	&	55396	&	-2.85E-03	&	-15.55 $\pm$ 2.8  	 \\
                	&		&	AE	&	114	&	50.2	&	57742	&	2.37E-03	&	12.901 $\pm$ 2.8  	 \\
	&		&		&		&		&		&		&		 \\
13292	&	1513	&	AB	&	100	&	17.9	&	27040	&	-0.0053	&	-38.23 $\pm$ 3.7  	 \\
	&		&	AC	&	164	&	11.8	&	17920	&	-0.0024	&	-16.99 $\pm$ 3.7  	 \\
	&		&	AD	&	133	&	23.1	&	34917	&	-0.0013	&	-9.19 $\pm$ 3.7	 \\
	&		&	AE	&	222	&	19.9	&	30075	&	0.0328	&	235.40 $\pm$ 3.7	 \\
	&		&	AF	&	222	&	32.2	&	48769	&	-0.0111	&	-79.91 $\pm$ 3.7	 \\
	&		&		&		&		&		&		&		 \\
16795	&	190	&	AB	&	356	&	0.8	&	148	&	-0.0036	&	-3.24 $\pm$ 0.5   	 \\
	&		&	AC	&	267	&	75.7	&	14374	&	0.0011	&	0.96 $\pm$ 0.5    	 \\
	&		&	AE	&	117	&	39.9	&	7589	&	-0.0237	&	-21.35 $\pm$ 0.5  	 \\
	&		&	AF	&	338	&	67.2	&	12768	&	0.0011	&	1.01  $\pm$ 0.5   	 \\

\hline

\end{tabular}
\label{tab:finaltrapeziatable}
\end{center}
\end{table*}


\section{Transverse Velocities and Masses of the Main Components}
\label{sec:transversevelocities}

\subsection{Transverse Velocities of the Main Components}
\label{subsec:transvel}

\begin{figure*}
\centering
\includegraphics[width=16.0cm,height=9.0cm]{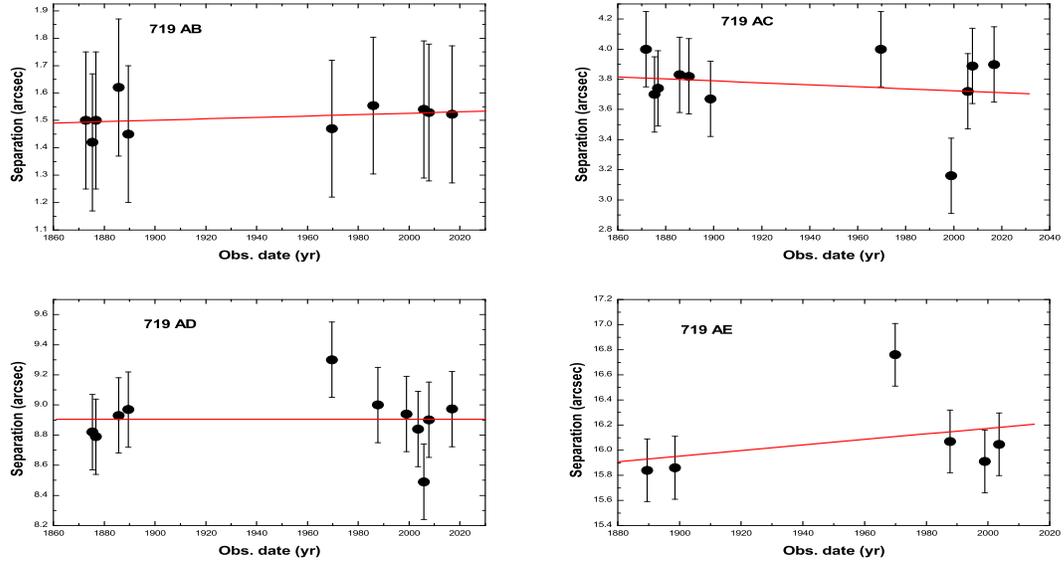}
\caption{Example of separation measurements with respect to the main component versus date of observation for the stars of trapezium ADS 719. The line represents the least squares fit to the points from which we obtained separation velocity in km/s.}
\label{fig:trap719_vel_figs}
\end{figure*}

\begin{figure*}
\centering
\includegraphics[width=16.0cm,height=9.0cm]{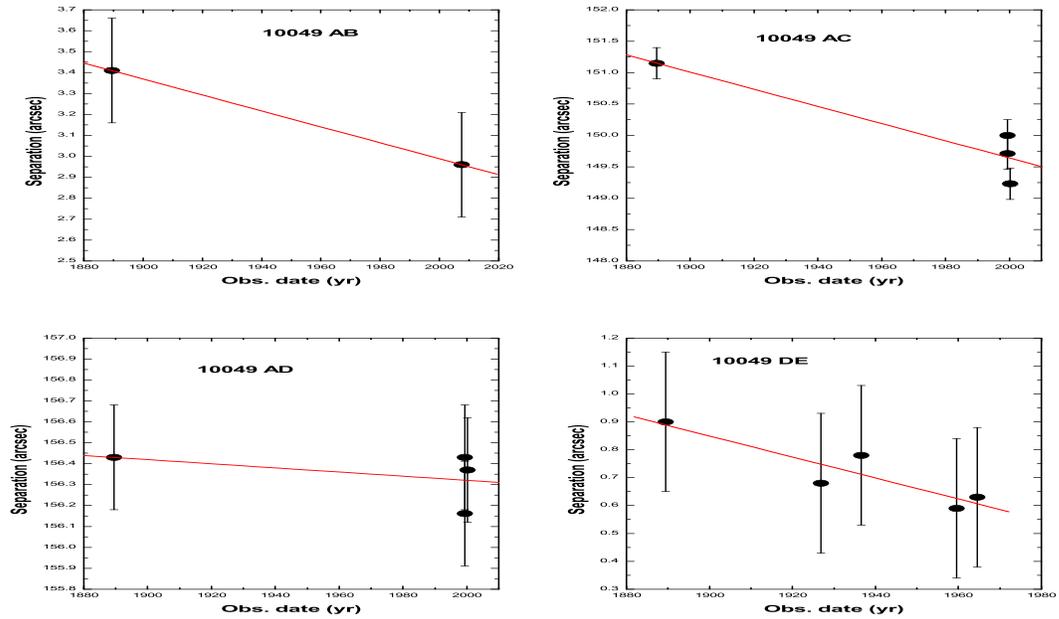}
\caption{Same as for Figure \ref{fig:trap719_vel_figs} but for ADS 10049.}
\label{fig:trap10049_vel_figs}
\end{figure*}

\begin{figure*}
\centering
\includegraphics[width=16.0cm,height=9.0cm]{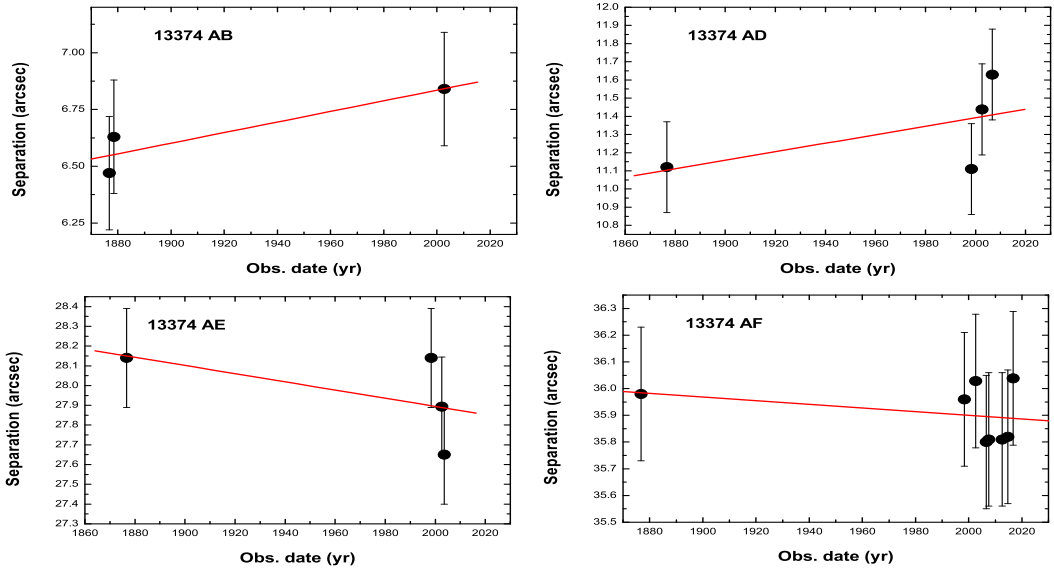}
\caption{Same as for Figure \ref{fig:trap719_vel_figs} but for ADS 13374.}
\label{fig:trap13374_vel_figs}
\end{figure*}

\begin{figure*}
\centering
\includegraphics[width=16.0cm,height=9.0cm]{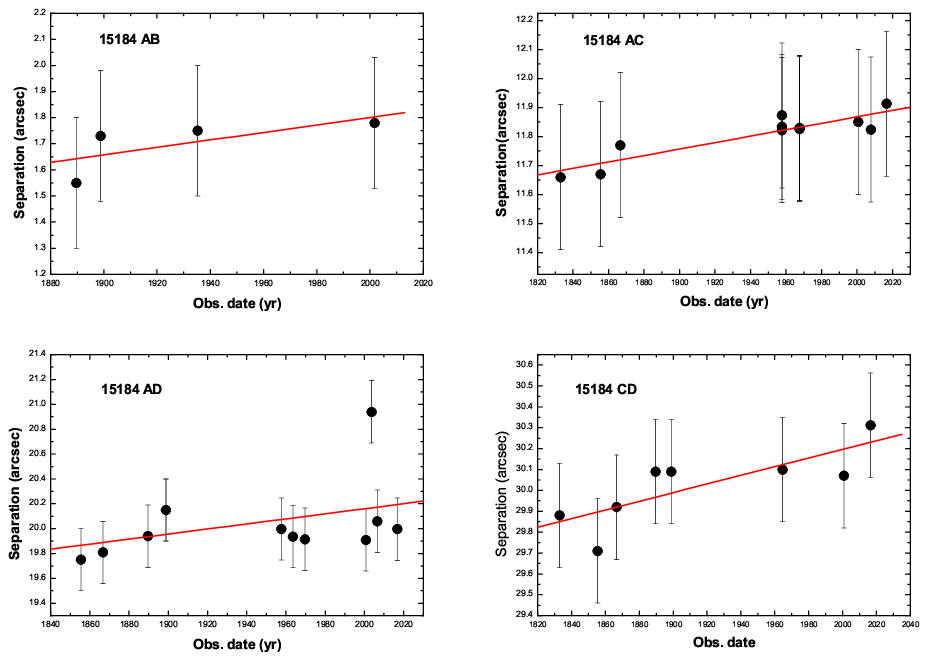}
\caption{Same as for Figure \ref{fig:trap719_vel_figs} but for ADS 15184.}
\label{fig:trap15184_vel_figs}
\end{figure*}

Using observations well separated in time for these systems we obtained the rate of change of separation between components in arcsec/yr. From the best available distance determinations to each one of the primaries, we calculated the components' transverse velocities (on the plane of the sky) in km/s. In addition, we obtained the X,Y positions with respect to the primary star for each component. In Figs. \ref{fig:trap719_vel_figs} to \ref{fig:trap15184_vel_figs} we show as examples the observed separations of the components of ADS 719, ADS 10049, ADS 13374 and ADS 15184 as a function of time. Note that the data for ADS 13374 and ADS 10049 are rather sparse.
{\bf We do not use proper motions because they are unavailable for all but a few of the secondary components. To estimate transverse velocities we use the relative motions among pairs of components.  Many of the data are historical. For the best observers, separations are usually accurate, but the position angles are very uncertain, as shown in \citet{Olivaresetal2013}.  The errors assumed for the historical visual observations were taken from \citet{Allen1974b}.  We only used measures by the ``Best" and ``Good" observers, as listed in \citet{Allen1974b}. Their errors are estimated to be from 0.07-0.08" for the ``Best" observers, and from 0.10-0.12" for the ``Good" observers. The errors for modern observations were estimated from the average differences in separation measures by the same observer at roughly the same time.  We are aware that by using only separation velocities we are underestimating the space velocities.  This would tend to make the systems more dynamically stable; as we shall see later, this is a conservative approach. }

Table \ref{tab:finaltrapeziatable} summarises the data available for the 10 selected trapezia. The first column gives the trapezium ADS number, the second the  distance in parsecs, the third shows the pairwise association with Component A, the primary trapezium star. {\bf Columns four, five, six and seven show the position angle (in degrees), the relative separation (in arcsec), the relative separation (in AU) and the rate of change of the separation of the  companion star with respect to star A (in arcsec per year). Finally, the last column lists the relative separation velocity (in kilometres per second)}.


\subsection{The Masses of the Main Components}
\label{subsec:masses}

The mass we assigned to each of the trapezium components was the one that corresponds to the spectral type of the component, when available. To this end, we adopted the empirical masses listed by \citet{Andersen1991} for early-type binary stars, those of \citet{HohleNeuhauserSchutz2010} for early-type stars and \citet{HabetsHeintze1981} for main sequence stars.  Although we also consulted more recent theoretical calibration we decided to use the empirical data, interpolating when necessary. We found a few cases with dynamical determinations of the mass of the primary (when it is a spectroscopic binary). Such cases were ADS 13374A, for which a dynamical mass was obtained by \citet{UnderhillHill2000}, ADS 15184A, with a dynamical mass obtained by \citet{HarvinGiesPenny2003}, ADS 11169A, for which we took the dynamical mass of \citet{HohleNeuhauserSchutz2010} and ADS 10049A, for which \citet{Novakovic2007} determined a dynamical mass. For the masses of the secondaries we proceeded as in the other cases. When no spectral type was available, mostly in the case of the weaker components, we drew upon the published B-V colour of these, corrected by the mean colour excess ---presumably due to interstellar reddening--- of the brighter components with published spectral type. This unreddened colour index was then assumed to be the intrinsic one of a main-sequence star to which a mass can be assigned according to the calibration by \citet{HabetsHeintze1981}.

Table \ref{tab:trapeziamassestable} summarises the adopted masses for the selected trapezia. We expect mass errors to be at least of the order of 10$\%$. {\bf If undetected companions are present among the stars studied, the masses could be much larger.  In fact, it is known that the majority of OB stars have close companions \citep{Sanaetal2012,Sanaetal2014,MoeDiStefano2017}.  To allow for these uncertainties, we will also study the effect of doubling the masses for some trapezium systems.}


\begin{table*}
\caption{Adopted masses for the main components of selected trapezia.}
\begin{center}

\begin{tabular}{ccccccc}

\hline

ADS		&	Component	& 	Mass (M$_\odot$) &		& ADS	&	Component	& 	Mass (M$_\odot$)	\\      
\hline        
		&	&	&	&	&	 &	\\
719		&	A	&	60	& 	&	11169	&	A	&	39	\\
		&	B	&	25	& 	&		&	B	&	 4	\\
		&	C	&	22	&	&		&	D	&	 8	\\
		&	D	&	19	& 	&		&	E	&	 7	\\
		&	&	&	&	&	&	\\
        
1869	&	A	&	12	& 	&	13292	&	A	&	12	\\		
		&	B	&	 6	& 	&	&	B	&	 4	\\
		&	C	&	 9	&	&	&	C	&	 3	\\
		&	&	&	&	&	&	\\
        
2843	&	A	&	12	& 	&	13374	&	A	&	31	\\
		&	B	&  2.5	&	&	&	B	&	 6	\\
		&	C	&  1.9	&	&	& 	C	&	12	\\
		&	D	&  2.5	&	&	&	D	&	10	\\
        &		&		& 	&	&	E	&	21	\\
        &	&	&	&	&	&	\\
4728	&	A	&	13	& 	&	15184	&	A	&	46	\\
		&	B	&	 9	&	&	&	B	&	18	\\
		&	C	&	13	&	&	&	C	&	10	\\
		&	D	&	 6	&	&	&	D	&	11	\\
        &	&	&	&	&	&	\\
10049	&	A	&	24	& 	&	16795	&	A	&	 6	\\
		&	B	&	 8	&	&	&	B	&	 2	\\
		&	C	&	 5	&	&	&	C	&	 4	\\
		&	D	&	12	&	&	&	E	&	 2	\\
		&		&		&	&	& 	F	&	 7	\\	
&	&	&	&	&	&	\\

\hline

\end{tabular}
\label{tab:trapeziamassestable}
\end{center}
\end{table*}


\section{The Dynamical Modelling}
\label{sec:dynamicalmodel}

\subsection{Initial Conditions and Numerical Integrations}
\label{subsec:init-con}

To model the dynamical evolution we conducted N-body simulations of the selected trapezium systems.  To accomplish this, we need a set of initial conditions, as well as masses for the individual components.  The  positions and {\bf transverse velocities}, along with the distance to each system, provide four of the needed initial conditions, as described in Section \ref{subsec:transvel}. The masses were assigned as described in the previous section. {\bf Radial velocities of the individual components are not readily available. In their study of trapezia, \citet{Abt1986} and \citet{AbtCor2000} obtained spectra of 120 stars in 31 candidate trapezia. They were able to classify these stars and estimate their distances.  However, their material had insufficient resolution to determine radial velocities.} 
We have no information about the extent in ``depth" of the systems. We adopt a Monte Carlo model to assign random values to these quantities, assuming that the $z$-coordinate will be contained within the radius of the system, and the $z$-velocity will be of the same order as the transverse velocities of the components. {\bf These quantities were drawn from normal distributions, centered respectively on zero and on the velocity centroid, and with  dispersions equal to half the radius and to the transversal velocity dispersion, respectively} . To the four known quantities we assign perturbations compatible with their observational uncertainties, as given in the previous sections. The perturbations are again drawn from normal distributions, with dispersions representing the uncertainties.   In this way, a set of 100 initial conditions was generated for each system, and these were the starting points for the numerical integrations.  
To integrate numerically these systems we used the code by \citet{Mikkola1993}, which implements a chain regularisation scheme and is well suited for small N-systems. Indeed, the fractional error in the energy at the end of the computations was always smaller than $10^{-11}$.
In the following paragraphs we present some of the results obtained for the ensembles associated with each system.


\subsection{The O-trapezia}
\label{subsec:otrap}

\begin{enumerate}

\item ADS 719
This 5-component system includes an O5V primary and thus resembles the Orion Trapezium, but it is situated at a much larger distance (1500 pc), and the component masses are smaller (if no undetected binaries are present). \citet{AbtCor2000} consider that component E is a background star. We ran simulations both including and excluding component E.  In the 5-component simulations, star E always escaped right at the beginning of the evolution. Also, a large fraction ($59\%$) of the generated systems turned out to be unbound, i.e. their total energy was positive. We obtained ``snapshots" of the evolutionary status at 1, 5, 10 and 25 crossing times.  A crossing time corresponds to about 10~000 years.
The 4-component simulations, which are surely more realistic, show that all generated systems are initially bound. Snapshots were obtained at 1, 2, 5, 10, 25, and 50 crossing times. The results are shown in Table \ref{tab:NumberSystems719}. Figure \ref{fig:coun719} displays as a function of time the number of different systems resulting from the evolution of the ensemble of 100 trapezia simulating ADS 719.  We show the numbers of trapezia of type (1,2,3,4), where all components remain at similar separations, and of type (1-2,3,4) where two components formed a close binary.  We also distinguish between hierarchical triples, HT, (a close binary attended by a distant companion, hence probably stable) and non-hierarchical triples, NHT, (three stars at similar separations, hence probably unstable). Finally, we show the number of binaries (B) formed.


\begin{table*}
\begin{center}
\caption{Number of different systems resulting at increasing crossing times for ADS 719.}
\begin{tabular}{cccccc}
\hline

Crossing Times	& 	Trapezia (t)	&	Trapezia (t)	&	Non-Hierarchical	&	Hierarchical	&	Binaries	\\
	            & 	(1,2,3,4)	    &  (1-2,3,4)	    &	Triples (NHT)	    &	Triples (HT)	&	(B)	        \\ 
\hline
                & 		            &		            &		                &		            &		        \\
0	            & 	100	            &	0	            &	0	                &	0	            &	0	        \\
1	            & 	46	            &	9	            &	30	                &	4	            &	11	        \\
2	            & 	36	            &	9	            &	43           	    &	4	            &	8	        \\
5	            & 	29	            &	5	            &	50	                &	6	            &	10	        \\
10	            & 	19	            &	8	            &	54	                &	3	            &	16       	\\
25	            & 	8	            &	4	            &	53	                &	10	            &	25	        \\
50	            & 	1	            &	3	            &	41	                &	19	            &	36	        \\
\hline

\end{tabular}
\label{tab:NumberSystems719}
\end{center}
\end{table*} 


\begin{figure}
\includegraphics[width=9.0cm,height=9.0cm]{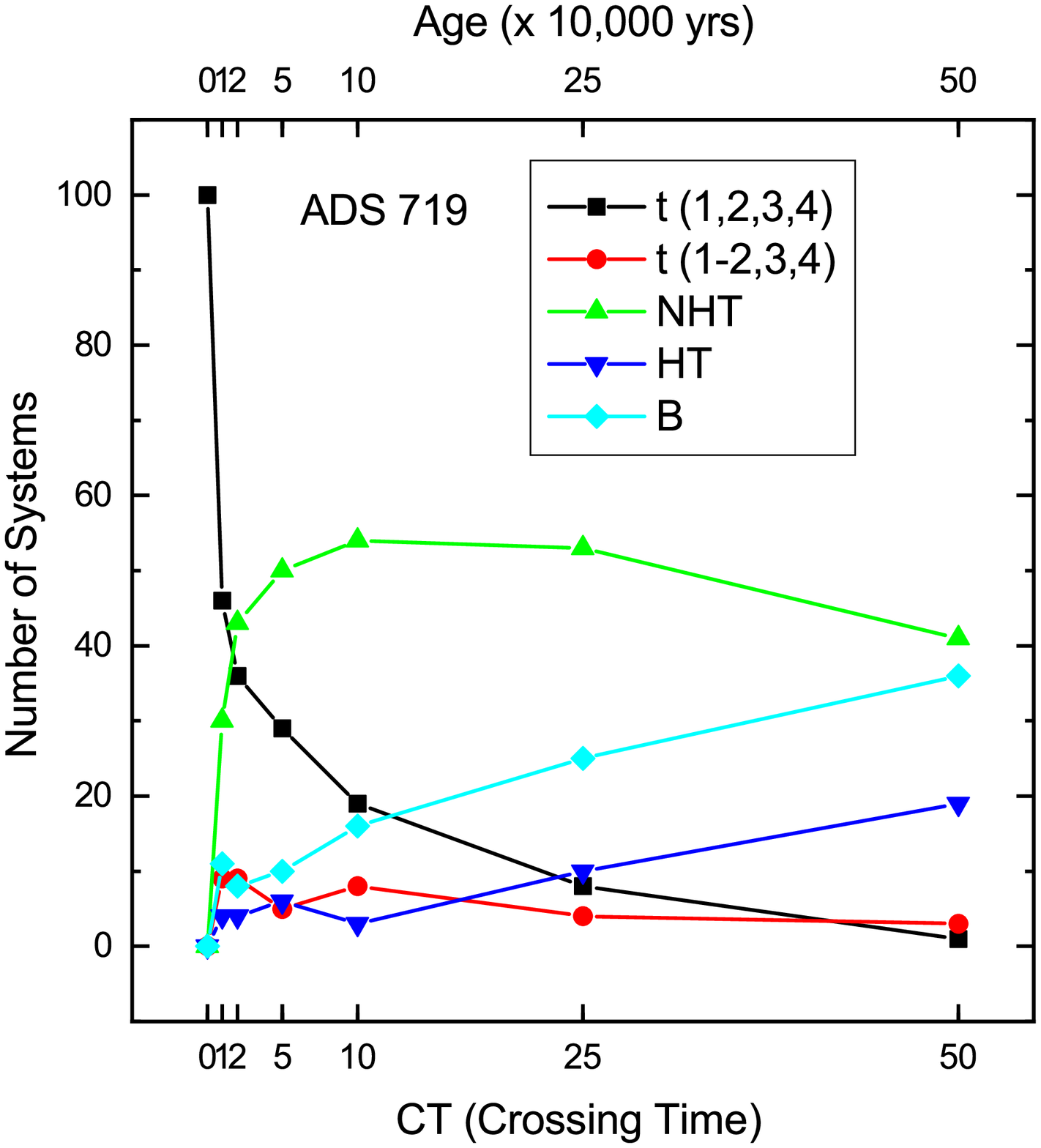}
\caption{Number of different systems generated from the simulations of an ensemble of 100 initial trapezia resembling ADS 719, as a function of crossing time or age.}
\label{fig:coun719}
\end{figure}


\item ADS 13374
This system comprises 6 components, with a WN5+O9I primary. However, component C has an extremely large transverse velocity (245 km/s), and is unlikely to belong to the trapezium. Thus, we excluded this component from the integrations. All the 5-component systems generated turned out to be unbound, and thus they rapidly disperse. Since the masses are quite uncertain due, among other things, to the presence of many undetected binaries and multiples among early-type stars, we ran simulations with twice the value of the masses. Even with this assumption the generated systems remained unbound. We also considered the possibility that the distance was overestimated, but we would have to assume ADS 13374 to be 10 times closer in order to obtain bound systems. This is certainly unrealistic. The results on this system, as well as other unbound ones, will be discussed later. Here we only note that ADS 13374 has the latest-type primary among the O trapezia, (WN5+O9I);  {\bf thus, the spectral type of the primary} already {\bf shows that it is an evolved star}. Note also the scarcity of historical positional data for this system.

\item ADS 15184
This trapezium has four components, with a primary of type O7V. Of the 100 generated systems, only two turned out to be bound. 
As in the case of ADS 13374 we ran simulations with twice the value for the masses. With this assumption, most (78) of the generated systems turned out to be bound. The dynamical evolution showed that after 1 crossing time 98 systems consisted only of a binary; the other two cases showed a bound hierarchical triple. After 10 crossing times only one system remained as a triple. Therefore, the dynamical evolution is in this sense complete after only 10~000 years. The dynamical lifetime of this trapezium, even assuming rather large values for the masses, is smaller than 10~000 years.

\end{enumerate}


\subsection{The B-trapezia}
\label{subsec:btrap}

\begin{enumerate}

\item ADS 1869
This trapezium has only 3 components, with a primary of type B2Vn.  All 100 generated systems turned out to be unbound, and they dispersed rapidly, within one crossing time.

\item ADS 2843
This trapezium has 5 components, with a primary of type B1I. However, Component D has a rather large transverse velocity (71 km/s) and is probably a non-member. We excluded it from the computations. The generated systems all result unbound, so they are already dissolved after the first crossing time. Even with twice the value for the masses the systems remain unbound. The distance (or the velocities) would have to be a factor of 10 smaller, for the the systems to become bound.  We consider this an unrealistic assumption.

\item ADS 4728
This is a 4-component trapezium, with a primary of type B1V. Just as in the case of ADS 2843, all the generated systems turned out to be unbound, and were already dispersed after the first crossing time. Again, we assumed twice the value for the masses, but all  the systems remained unbound.  

\item ADS 10049
This trapezium has four components with a primary of type B2V.  Only 32 out of 100 simulated systems resulted bound. The unbound systems quickly dissolved, often producing a binary. After 10 crossing times the bound systems consisted of a triple.  Star C (the least massive one) always escaped right at the beginning of he runs.  The lifetime of this system is thus less than 10~000 yr.
Assuming masses twice as large we obtained bound systems in 95 of the generated cases. The dynamical evolution showed that the least massive star (Component C) almost always escaped, leaving a triple system. A strong tendency for Components A and B to form a binary is observed (in 97 of the cases). Results at 1, 2, 5 and 10 crossing times are shown in Table \ref{tab:NumberSystems10049} and Figure \ref{fig:count10049} for the larger mass values. Although the lifetime of this system as a 4-component trapezium is less than 10~000 yr, the non-hierarchical (and thus unstable) triple systems survived for longer times;  after 10 crossing times (100~000 yr) there were still 13 non-hierarchical triples present.  


\begin{table*}
\begin{center}
\caption{Number of systems resulting at increasing values of the crossing time for ADS 10049 taking the masses at twice their nominal values.}
\begin{tabular}{cccccc}
\hline

Crossing Times	& 	Trapezia (t)	&	Trapezia (t)	&	Non-Hierarchical	&	Hierarchical	&	Binaries	\\
	            & 	(1,2,3,4)	    &  (1-2,3,4)	    &	Triples (NHT)	    &	Triples (HT)	&	(B)	        \\ 
\hline
                & 		            &		            &		                &		            &		        \\
                       0	            & 	100          	&	0	            &	0	                &	0	            &    0	     \\
                        1	            & 	0	            &	0	            &	8	                &	79	            &	13	     \\
                       2           	& 	0	            &	0	            &	12	                &	75	            &	13	      \\
                       5	            & 	0	            &	0	            &	16	                &	65	            &	19	      \\
                     10          	& 	0	            &	0	            &	13	                &	56	            &	31        \\

\hline

\end{tabular}
\label{tab:NumberSystems10049}
\end{center}
\end{table*} 


\begin{figure}
\includegraphics[width=9.0cm,height=9.0cm]{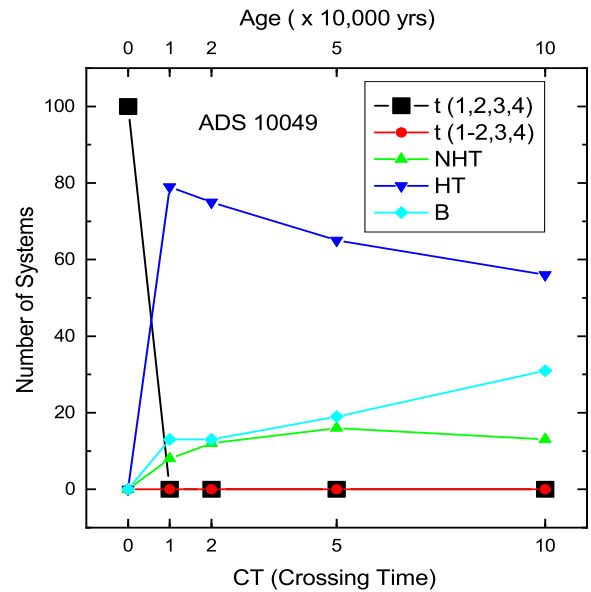}
\caption{Number of different systems generated from 100 initial trapezia simulations for ADS 10049, as a function of crossing time or age. The simulation assumes twice the values for the adopted masses.}
\label{fig:count10049}
\end{figure}


\item ADS 11169
Five components are listed for this system, with a primary of type B9Ia. However, Component C is likely to be {\bf a foreground star} according to \citet{AbtCor2000}. Indeed, it has a large transverse velocity (27 km/s), and was excluded from the integrations. All generated 4-component systems turned out to be unbound, and are already dispersed after 1 crossing time. Even assuming masses a factor of two larger, the generated systems remain unbound.  

\item ADS 13292
Here we have a 3-component system, with a primary of type B1Vn.  Just as for the case of ADS 1869, all 100 generated systems turned out to be unbound, and they dissolved within one crossing time.  

\item ADS 16795
This trapezium has five components, with a primary of type B3V.  The generated systems all result unbound, and quickly disperse, in less than one crossing time. The dynamical status of this trapezium is similar to that of ADS 1169, ADS 2843 and ADS 11169, which are all unbound. They will be further discussed below.

\end{enumerate}


\section{Results}
\label{sec:results}

\subsection{The Number of Unbound Systems}
\label{subsec:unbound}

The number of unbound systems found is surprisingly large. Indeed, taking values for the masses as in Table \ref{tab:trapeziamassestable}, 9 out of 10 systems turned out to be unbound. This is strange, since the systems were carefully selected to exclude optical companions and pseudo-trapezia. Of course, one expects star formation to occur in bound protostellar entities, and trapezia are young systems, which are expected to continue to be bound.  The large number of unbound systems found among the modelled ensembles of trapezia could be due to the presence of one (or more) optical companions in the system.  An illustrative example of this situation is ADS 719. This trapezium has 5 components, one of which, Component E, is considered to be a optical by \cite{Abt1986}. A test ensemble was generated including Component E, and resulted in 69 unbound systems.  If Component E is excluded, as it should be, the generated ensemble resulted in 100 bound systems. A similar situation could be happening in other cases, in which one (or more) unidentified optical companion has been inadvertently included in the modelled ensembles.  Another possibility is that, as a result of a very early dynamical evolution, one (or more) binaries have already formed, but are as yet undetected. These early binaries would absorb a large fraction of the binding energy of the primeval trapezium, and could account for the apparent unboundedness of the systems. With  average major semi-axes of 2000 AU these undetected binaries, situated at distances of roughly 1000 pc (or more) would show separations of 2 arcsec or less, with vanishingly small traces of orbital motion. They would show small common proper motions and common radial velocities. The discovery of these binaries, resulting from the early dynamical disintegration of trapezia, is thus a challenging problem for observers. Yet another possibility could be a massive gas loss in the very early stages of formation. This has been recently proposed by \cite{Bravietal2018} to explain the supervirial status of three out of four open clusters they studied using $Gaia$-ESO Survey data. 

We carried a test of our conjecture that the observed trapezia have already undergone considerable dynamical evolution, that is, that they are dynamically ``old", by looking for HII regions around the O-type trapezia. We examined the red images in 2MASS and SDSS. If HII regions are similar to the prototype, the Orion Nebula, they are expected to dissipate after about 15~000 yr \citep{ODell2009}. Of the three O-type trapezia in our list, only ADS 719 is still surrounded by a bright nebula.  We interpret this as an indication that ADS 13374  and ADS 15184 are now already dynamically ``old", older than 15 000 years, the estimated age of the Orion Nebula \citep{ODell2009}.  This time is sufficient for substantial dynamical evolution \citep{Allen2015, Allen2017}.  We note that among the O-type trapezia, ADS 719 has the earliest spectral type, and could be the youngest. In contrast, ADS 13374 has the latest spectral type and, {\bf having a primary of spectral type WN5+O9I}, already shows signs of stellar evolution.  This is entirely consistent with the dynamical results we find. As we discuss below, the dynamical lifetime of ADS 719 is 10~000 to 20~000 yr, while ADS 13374 appears now as an unbound system, without traces of an HII region around it, presumably because it is much older than 15~000 years, has already dispersed its nebula, and has undergone significant dynamical evolution.


\subsection{The Lifetimes of the Modelled Systems}
\label{subsec:lifetimes}

For a quantitative assessment of the lifetimes, we define the ``dynamical lifetime" of a system as the time necessary for one half of the original population to have evolved into non-trapezium configurations.  
The lifetimes of the unbound systems are, of course, extremely short, of the order of a crossing time (10~000 yr) or less.  The unbound systems are ADS 2843, ADS 13374, ADS 11169, ADS 16795, ADS 15184 (with nominal values for the masses of the components), ADS 10049 (also with nominal masses), ADS 1869, ADS 13292 and ADS 4728. But, as discussed above, it makes little sense to talk about lifetimes of unbound systems, since they could include optical companions, or could be already in a later stage of their dynamical evolution, with one or more undetected binaries.  

The lifetimes of the bound systems are also quite short. In Table \ref{tab:NumberSystems719} and Figure \ref{fig:coun719} we present the results for ADS 719. It can be seen that after only 2 crossing times (20~000 yr) only 41 trapezia remained as such, out of the original 100.  So, the lifetime of ADS 719 as a four-component trapezium is between 10~000 and 20~000 yr. However, at TC=2 a further 43 non-hierarchical triples remained, and even at TC = 25 still 53 non-hierarchical triples survived. So, the lifetime of the dynamically unstable subsystems remaining from the evolution of ADS 719 is longer, of the order of 250~000 -- 500~000 yr. 


The trapezium ADS 10049 with the masses as given in Table \ref{tab:trapeziamassestable} turned out to be unbound in 68 percent of the cases, with a lifetime much shorter than 10 000 yr. Star 3 escaped right at the beginning of the runs. We decided to simulate this system assuming the extreme case that it was composed of undetected binaries, and took masses  twice as large for the individual components.  With this assumption 94 percent of the systems resulted bound.  We obtained the results shown in Table \ref{tab:NumberSystems10049} and Figure \ref{fig:count10049}. Already after 1 crossing time (10~000 yr) the systems disintegrated, giving mostly hierarchical triples.  In 97 percent of the cases Stars A and B quickly formed a binary, accompanied by a distant Star D. In 97 percent of the cases Star C was an early escaper.  The lifetime of this system, even assuming masses twice as large as given in Table \ref{tab:trapeziamassestable}, is thus less than 10~000 yr.
{\bf The very short lifetimes we find, as well as the large number of unbound systems, indicates that trapezia should be rare. Actually, we found only 3 true O trapezia.  The number of O stars known with B$<$12 \citep{MaizApellanizetal2013} is about 800. The mean age of an O star is about 500~000 years, which is large compared to  the dynamical lifetimes of bound trapezia (at most 50~000 years for the Orion Trapezium, \citet{Allen2017}). If O stars are formed in trapezium-like groups (it appears difficult to form isolated massive stars), then roughly 80 O-trapezia should be found. The scarcity of observed true trapezia is suggestive of very short dynamical lifetimes for these systems.  Only a few survivors are actually found.}

\subsection{The Velocity Distribution of the Ejected Stars}
\label{subsec:ejectedstars}

\begin{figure*}
\includegraphics[width=20.0cm,height=20.0cm]{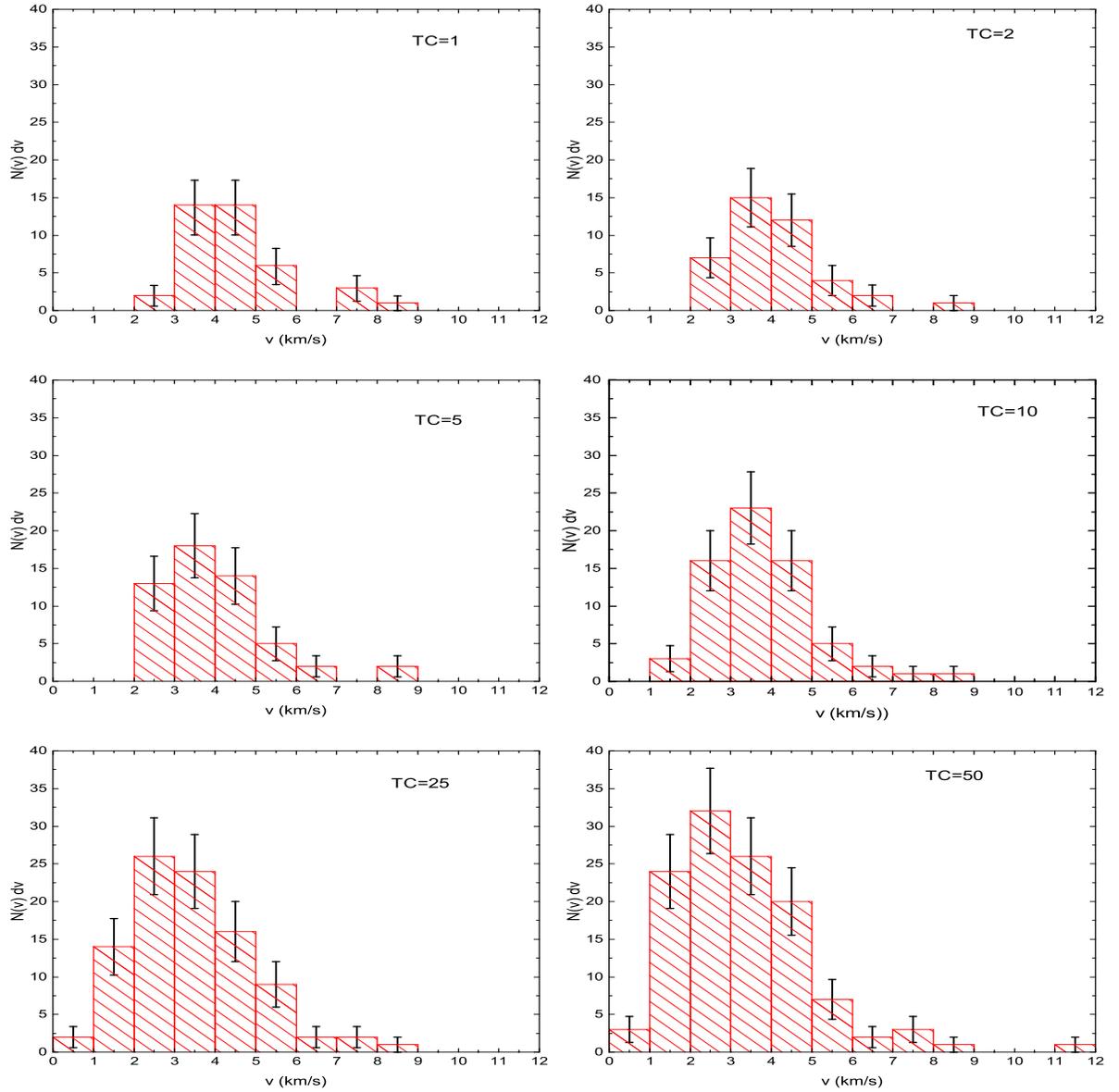}
\caption{Frequency distribution of the velocities of the ejected stars for ADS 719. Results are shown after 1, 2, 5, 10, 25 and 50 crossing times. The error bars correspond to  $\pm ~ \sqrt{N}$.}
\label{fig:hist719escaped}
\end{figure*}

Figure \ref{fig:hist719escaped} shows, as an example, the velocity distribution of the stars that escaped during the simulation of ADS 719. As the figure shows, the ejected stars have small velocities, of the order of a few km/s. In rare cases, velocities as large as 10 km/s appeared, signaling a few close encounters. Simulations for the other trapezia also produced low velocity escapers, similar to the results for ADS 719. No runaway stars were formed.


\subsection{The Properties of the Binaries Formed in the Numerical Simulations}
\label{subsec:binariesformed}
\begin{figure*}
\includegraphics[width=20.0cm,height=20.0cm]{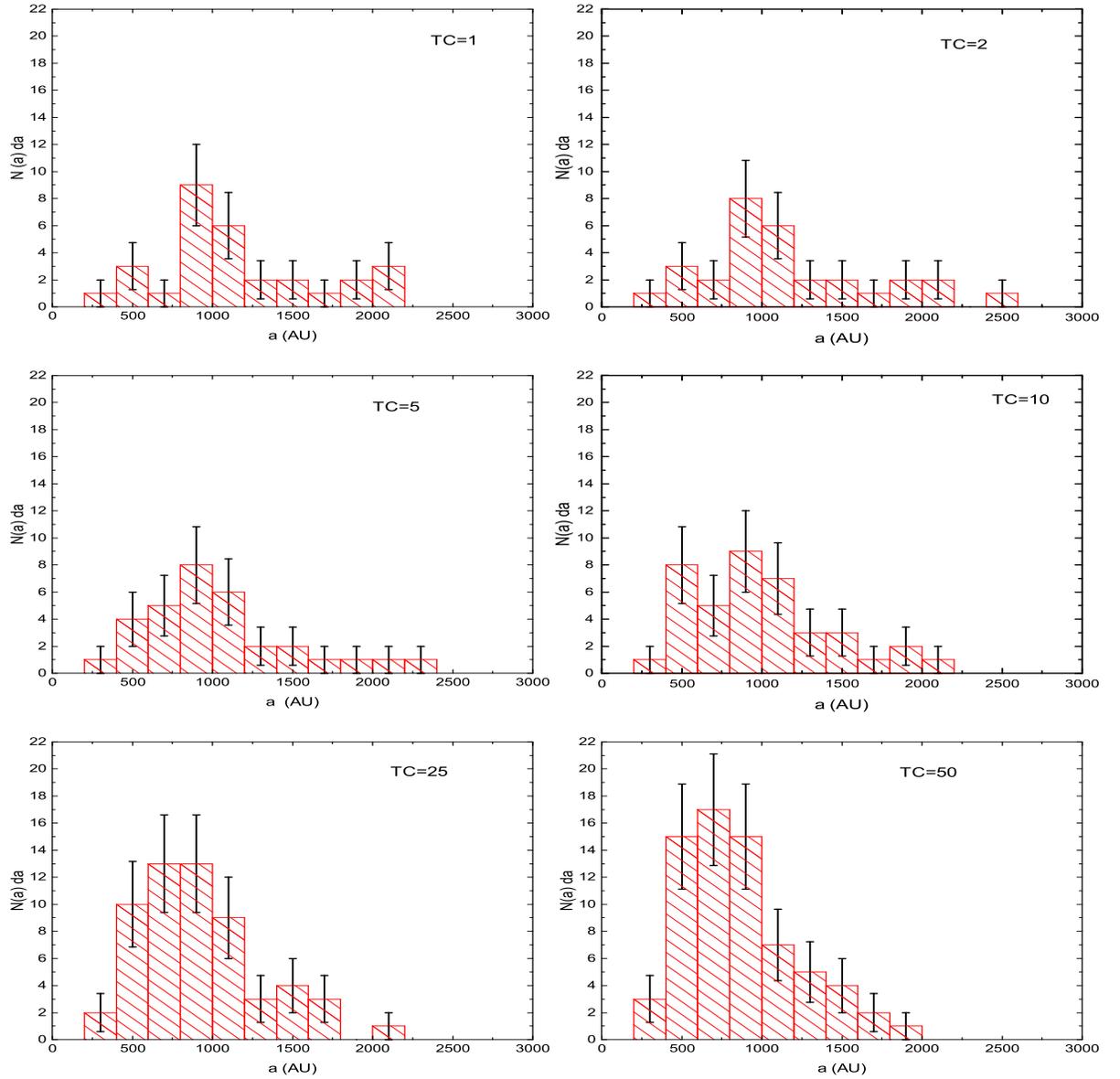}
\caption{Frequency distribution of the semi-axes of the binaries formed for ADS 719. Results are shown after 1, 2, 5, 10, 25 and 50 crossing times. The error bars correspond to  $\pm ~ \sqrt{N}$.}
\label{fig:hist719binaries}
\end{figure*}

\begin{figure*}
\includegraphics[width=20.0cm,height=20.0cm]{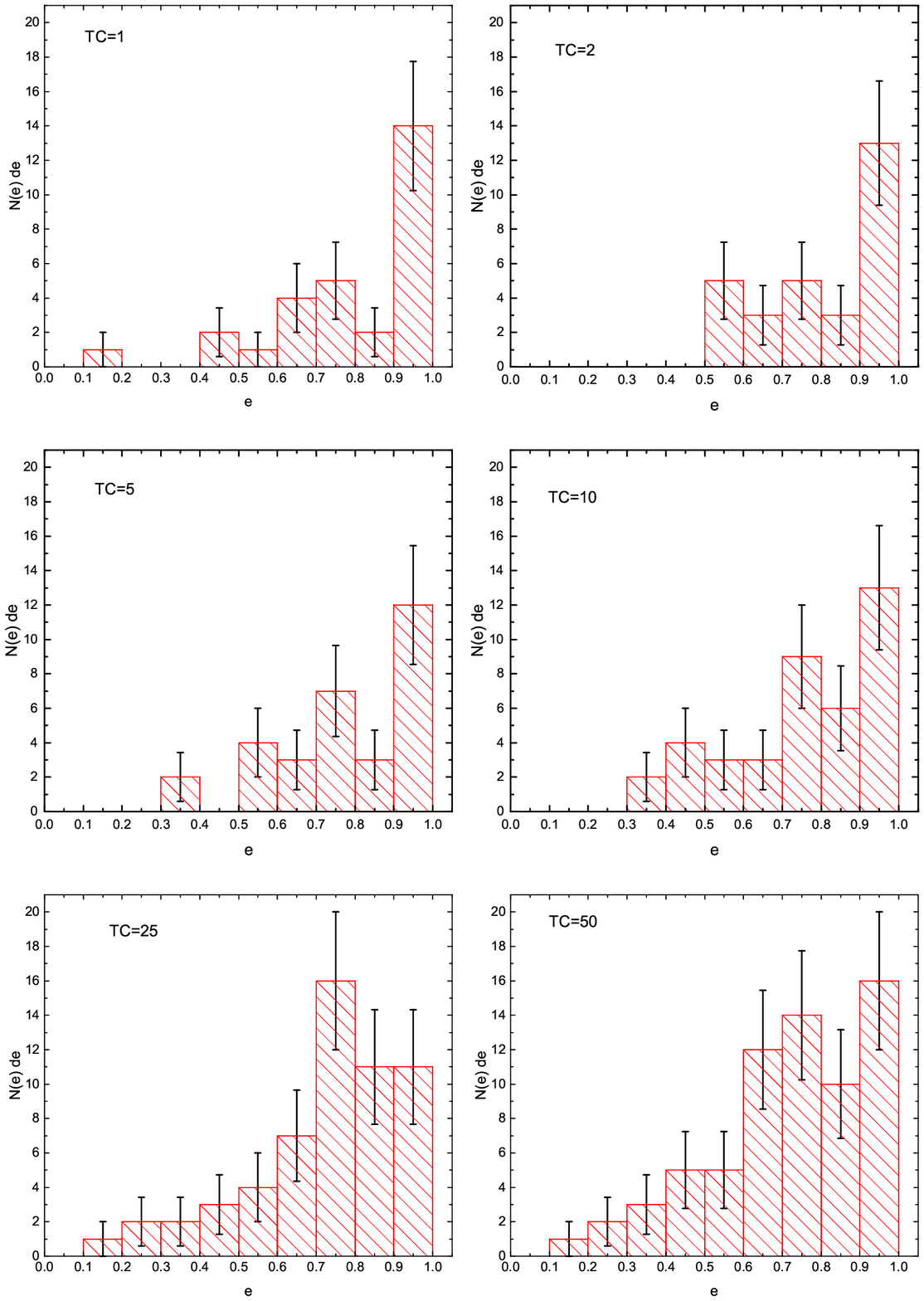}
\caption{Frequency distribution of the eccentricities of the binaries formed for ADS 719. Results are shown after 1, 2, 5, 10, 25 and 50 crossing times. The error bars correspond to  $\pm ~ \sqrt{N}$.}
\label{fig:hist719excentricities}
\end{figure*}

\begin{figure}
\centering
\includegraphics[width=9.0cm,height=9.0cm]{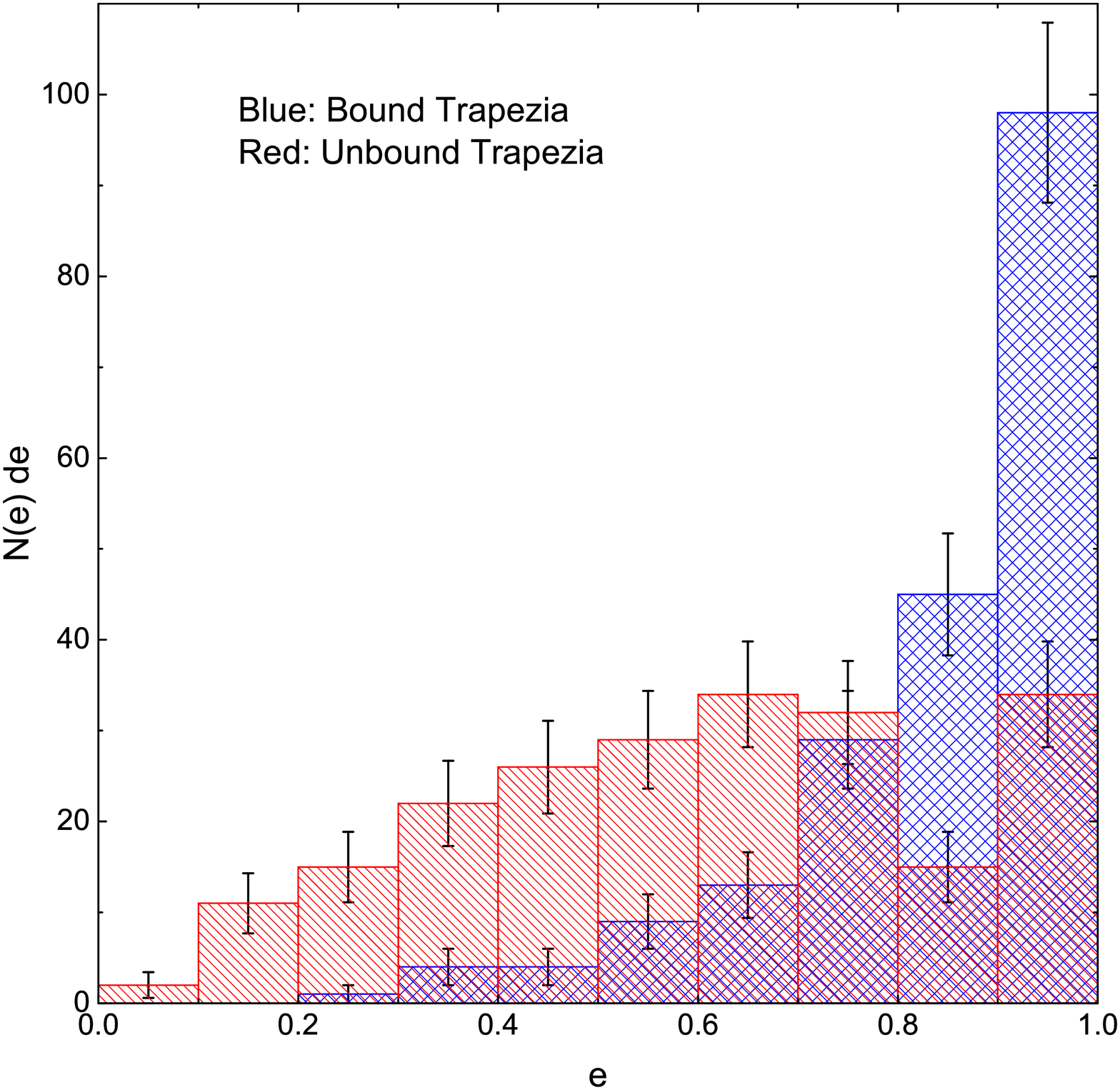}
\caption{Combined frequency distribution of the eccentricities of binaries formed in bound (blue cross-hatched) and unbound (red shaded) trapezia.}
\label{fig:combinedexcentr}
\end{figure}

\begin{figure}
\centering
\includegraphics[width=9.0cm,height=9.0cm]{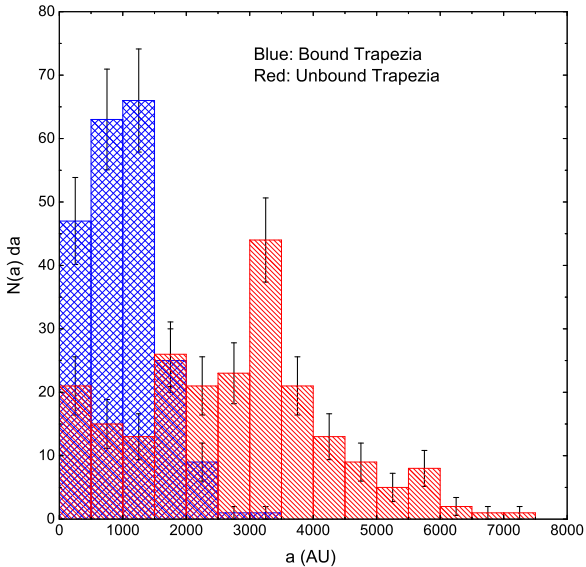}
\caption{Combined frequency distribution of the major semi-axes of binaries formed in bound (blue cross-hatched) and unbound (red shaded) trapezia.}
\label{fig:combinedsemiaxes}
\end{figure}

\begin{figure}
\centering
\includegraphics[width=9.0cm,height=9.0cm]{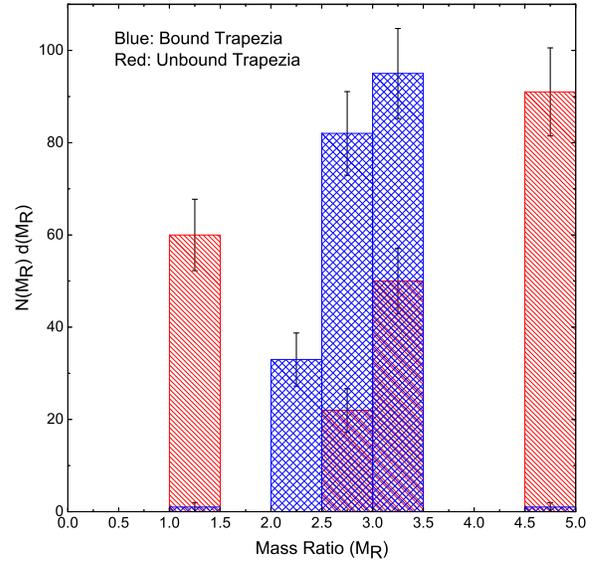}
\caption{Combined frequency distribution of the Mass ratios for binaries formed in bound (blue cross-hatched) and unbound (red shaded) trapezia.}
\label{fig:cocientemasas}
\end{figure}

\begin{figure}
\includegraphics[width=9.0cm,height=9.0cm]{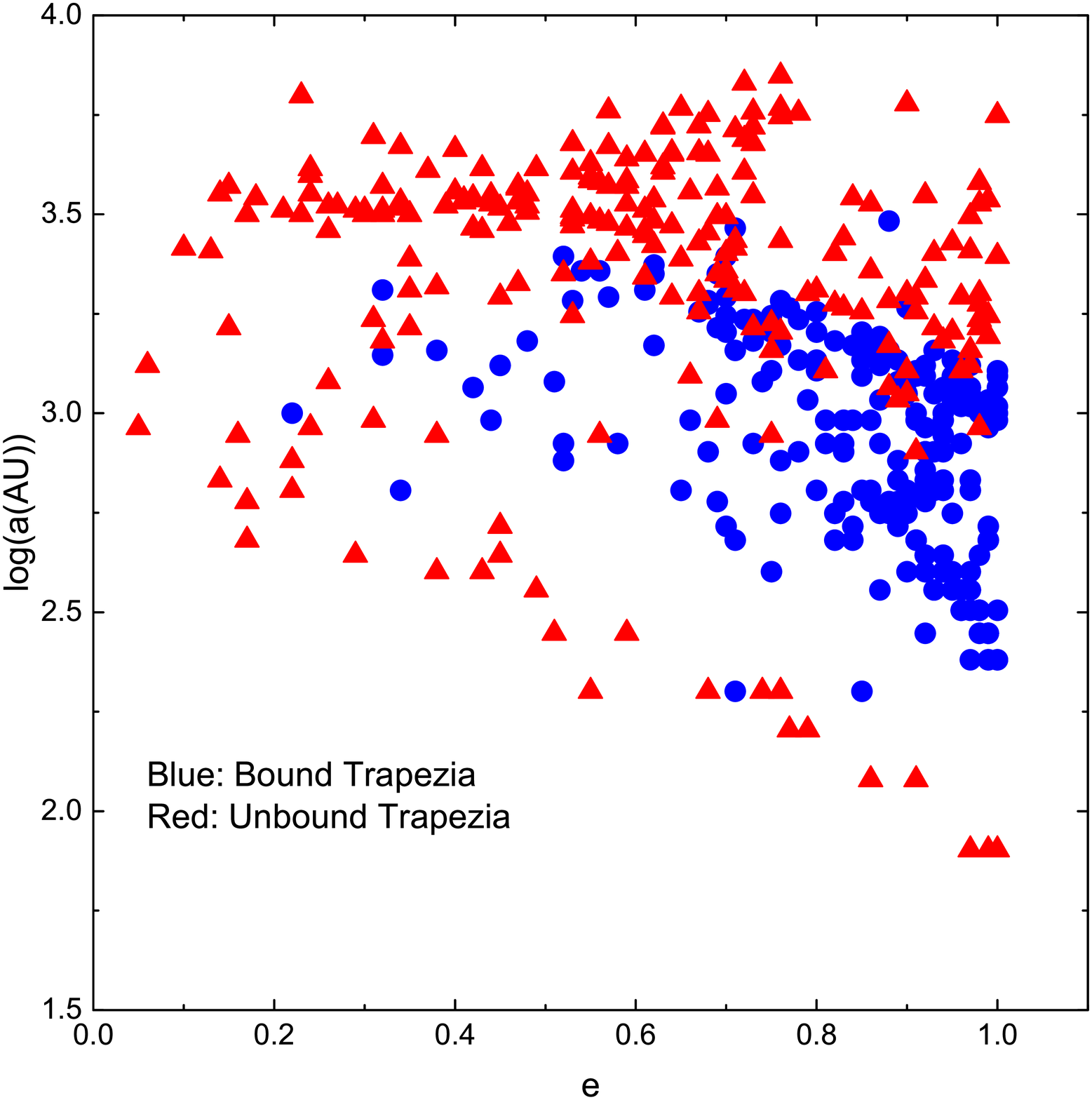}
\caption{Major semi-axes as a function of eccentricities for all the binaries formed in the numerical integrations. Very few binaries with eccentricities smaller than 0.2 are present among those formed in bound systems, a result in accord with observations. Dots correspond to bound systems and triangles to unbound systems.}
\label{fig:semiaxesvsexcen}
\end{figure}

As an example, Figure \ref{fig:hist719binaries} shows the distribution of major semi-axes of the binaries formed during the evolution of the ensembles resembling ADS 719, Figure \ref{fig:hist719excentricities} that of the eccentricities. The major semi-axes cluster around a value of 750 AU.  
The bound systems produced binaries as a result of their dynamical evolution.  Some of the unbound systems also produced binaries, very early in the computations. All these binaries will presumably end up as field binaries, and we show their combined properties in Figures \ref{fig:combinedexcentr}, \ref{fig:combinedsemiaxes} and \ref{fig:cocientemasas}. 
The properties of the binaries for the bound and unbound systems appear to be different (see Figs. \ref{fig:combinedexcentr}, \ref{fig:combinedsemiaxes} and \ref{fig:cocientemasas}). The orbital eccentricities of the binaries coming from both bound and unbound systems tend to a thermal distribution \citep{Heggie1975}, but the binaries produced in the unbound systems show a considerably flatter distribution. The mean values for bound and unbound systems are also different, $\langle$e$\rangle$ = 0.81 and $\langle$e$\rangle$ = 0.62, respectively. For comparison, the mean value expected for a thermal distribution is $\langle$e$\rangle$ = 0.66 \citep{Ambartsumian1937}.  Many binaries stemming from unbound systems have eccentricities smaller than ($\sim 0.5$), and as low as ($\sim 0.05$), whereas for the bound systems no binaries with eccentricities of less than  ($\sim 0.35-0.45$) are found. Observationally, there is no consensus on the distribution of eccentricities of field binaries. While \cite{DuquennoyMayor1991} {\bf as well as \citet{MoeDiStefano2017} find that it is thermal (at least for the larger semiaxes)}, \cite{DucheneKraus2013} and \cite{Raghavanetal2010} do not. 
As far as the values of major semi-axes are concerned, the bound systems favour the formation of binary systems with values of $a$ with a maximum around $\sim $ 750 AU. This value corresponds approximately to the situation where the binding energy of the initial system is absorbed by a single binary.  
As can be expected, the unbound systems favour the formation of wider binaries. The distribution shows a maximum at $\sim$  3500--4000 AU, but many wider binaries are produced, with semi-axes as large as 20 000 AU.  {\bf However, most of the binaries with $a$ $>$ 7000 AU turned out to be transitory associations and did not persist over time.  They were not plotted in Figures 10-13.} 
Although field binaries with $a$ values similar to those produced in the bound systems are frequent, the distribution of major semi-axes found in our simulations is not directly comparable with that of the observed field binaries, since they presumably stem from different formation processes.  Even if most of them originate in trapezium-type systems their binding energies (and hence the semi-axes of the binaries formed) will be different. The observed distribution of semi-axes is thus likely to be a superposition of the binding energies of many dissolved systems. 
{\bf Figure \ref{fig:cocientemasas} presents the distribution of the mass ratios (large mass/small mass) for the binaries formed in bound (blue cross-hatched) and unbound (red shaded) trapezia. The distributions are different.  Whereas binaries from bound trapezia have mass ratios in the range 2.0 to 3.5, those from unbound trapezia tend to more extreme values, either towards lower (1.0-1.5) or larger values (4.5-5.0). }
Figure \ref{fig:semiaxesvsexcen} is a plot of the major semi-axes as a function of eccentricity for all simulated ensembles. The red dots, corresponding to the bound systems show that very few systems with eccentricities smaller than 0.2 are found, at all values of the semi-axes.  This is in accordance with observations \citep{Raghavanetal2010, Sanaetal2012, Sanaetal2014, MoeDiStefano2017}. For larger values of the eccentricities the distribution seems to be random. The figure also shows that unbound systems produce binaries with more extreme characteristics. For semi-axes smaller than about 5~000 AU most binaries have eccentricities smaller than 0.4, and quite a few appear at e $\le$ 0.2. This does not accord with observations. Wider binaries appear at values of the eccentricity mostly larger than 0.5 and reaching up to 0.99. From this figure we can conclude that binaries stemming from bound systems are in better agreement with observations than those formed in unbound systems.


\section{Summary and Conclusions}
\label{sec:conclusions}

Our main conclusions can be summarised as follows.

\begin{enumerate}
\item After an extensive search of literature data we ended up with only 10 candidates to physical trapezia, 3 of spectral type O and seven of type B3 and earlier.

\item The available data included distances, masses for the primaries, and transversal velocities for the components.

\item Dynamical masses for the primaries (when available) were taken.  Otherwise they were assigned from their spectral types, when available, or from their colours, appropriately unreddened. Masses for the secondaries were assigned from their unreddened colours.

\item Numerical N-body simulations were conducted for ensembles of 100 systems resembling each one of the trapezia. Radial velocities of a magnitude similar to the transversal velocities were randomly assigned.

\item Most of the systems turned out to be unbound. Possible reasons include the inadvertent presence of unidentified optical companions, and an already dynamically evolved (``old") status for the systems, with one or more binaries already formed.

\item The lifetimes of the bound systems are short, between 10~000 and 20~000 yr. The unbound systems disintegrate in less than 10~000 yr. However, a large number of non-hierarchical --and hence unstable-- triples survive even at 50~000 yr. {\bf The small numbers of trapezia found among the known O stars is suggestive of very short lifetimes for these systems.}

\item The ejected stars show a velocity distribution with a shallow maximum around  2.5--3.5 km/s. They are low velocity escapers in all cases, that is, no runaway stars are formed.

\item A large number of binaries was formed during the evolution of both the bound and the unbound systems. The properties of these binaries, which would presumably end up as being part of the observed field wide binaries, show major semi-axes of a few hundreds to thousands AU, and eccentricities tending to thermal. Binaries from non-bound systems tend to have nearly circular orbits, those from bound systems show higher eccentricities. The properties of binaries formed in bound systems agree better with observations than those formed in unbound systems.

\end{enumerate}

{\bf As stated in Section 2 above, the recent publication of the Second $Gaia$ Data Release should be potentially important for studies of the kinematical and dynamical behaviour of trapezium systems. However, there are still a number of potential problems related to the treatment of close companions in Gaia, the possible presence of many undetected binaries, and the lack of radial velocities for early type stars. When improved data become available, this study, as well as our previous studies on the dynamical evolution of the Orion Trapezium \citep{Allen2017} and of the minicluster Theta 1 Ori B \citep{Allen2015} should be repeated.}


\vskip1.0cm


\vskip1.0cm


\section*{Acknowledgments}

We thank the support provided by Instituto de Astronom\'{\i}a at Universidad Nacional Aut\'onoma de M\'exico (UNAM). We also thank Direcci\'on General de Asuntos del Personal Acad\'emico, DGAPA at UNAM for financial support under projects number PAPIIT IN102517 and IN102617. We would like to thank Juan Carlos Yustis for his help with the figures presented in this paper. This work has made use of the Washington Double Star Catalogue maintained at the U.S. Naval Observatory. {\bf The constructive remarks of an anonymous referee are gratefully acknowledged.}

\vskip1.0cm



\label{lastpage}


\begin{thebibliography}{}


\bibitem[Abt~(1986)]{Abt1986} Abt, H.A., 1986, ApJ, 304, 688 

\bibitem[Abt \& Corbally~(2000)]{AbtCor2000} Abt, H.A. \& Corbally, C. J., 2000, ApJ, 541, 841

\bibitem[Allen \& Poveda~(1974)]{Allen1974a}Allen, C. \& Poveda, A. 1974, IAU Symposium, 62, 239

\bibitem[Allen et al.~(1974)]{Allen1974b}Allen, C., Poveda, A. \& Worley, C.~E. 1974, RMxAA, 1, 101

\bibitem[Allen et al.~(2004)]{Allen2004}Allen, C., Poveda, A. \& Hern\'andez-Alc\'antara, A. 2004, RMxAC, 21, 195

\bibitem[Allen et al.~(2015)]{Allen2015}Allen, C., Costero, R. \& Hern\'andez, M. 2015, AJ, 150, 167

\bibitem[Allen et al.~(2017)]{Allen2017}Allen, C., Costero, R., Ruelas-Mayorga, A. \&  S\'anchez. L.~J. 2017, MNRAS, 466, 4937

\bibitem[Ambartsumian~(1937)]{Ambartsumian1937} Ambartsumian, V.~A. 1937, Astron. J. USSR, 14, 207

\bibitem[Ambartsumian~(1954)]{Ambartsumian1954} Ambartsumian, V.~A. 1954, LIACo, 5, 293

\bibitem[Andersen~(1991)]{Andersen1991} Andersen, J. 1991, Astron Astrophys Rev., 3, 91

\bibitem[Bravi et al.~(2018)]{Bravietal2018} Bravi, L., Zari, E., Sacco, G.G. et al., 2018, ArXiv e-prints [arXiv:1803.01908] 

\bibitem[Duch{\^e}ne \& Kraus (2013)]{DucheneKraus2013}Duch{\^e}ne, G. \& Kraus, A., 2013, ARAA, 51, 269   

\bibitem[Duquennoy \& Mayor (1991)]{DuquennoyMayor1991}Duquennoy, A. \& Mayor, M., 1991, AA, 248, 485 
 
\bibitem[Habets \& Heintze~(1981)]{HabetsHeintze1981} Habetz, G.~M.~H.~J. \& Heintze, J.~R.~W., 1981, A\&AS, 46, 193

\bibitem[Harvin, Gies \& Penny~(2003)]{HarvinGiesPenny2003} Harvin, J.~A., Gies, D.~R. \& Penny, L.~R. 2003, Bull. Amer. Astron. Soc., 35, 1223

\bibitem[Heggie~(1975)]{Heggie1975}Heggie, D.~C. 1975, MNRAS, 173, 729

\bibitem[Hohle, M.~M., Neuh{\"a}user, R. \& Schutz, B.~F.~(2010)]{HohleNeuhauserSchutz2010} Hohle, M.~M., Neuh{\"a}user, R. \& Schutz, B.~F. 2010, Astron. Nach., 331, 349 

\bibitem[Ma\'iz Apell\'aniz et al.~(2013)]{MaizApellanizetal2013}Ma\'iz Apell\'aniz, J., Sota, A., Morrell, N.~I., et al. 2013, {\it Massive Stars: From $\alpha$ to $\Omega$, Rhodes Greece, 10-14 June 2013}

\bibitem[Mason et al.~(2001)]{Mason2001}Mason, B.~D., Wycoff, G.~L., Hartkopf, W.~I., et al. 2001, AJ, 122, 3466

\bibitem[Mikkola \& Aarseth~(1993)]{Mikkola1993} Mikkola, S. \& Aarseth, S. 1993, CeMDA, 57, 439

\bibitem[Moe \& Di~Stefano~(2017)]{MoeDiStefano2017} Moe, M. \& Di~Stefano, R. 2017, ApJS, 230, 15

\bibitem[Novakovi{\'c}~(2007)]{Novakovic2007} Novacovi{\'c} , B. 2007, Baltic Astron., 16, 435

\bibitem[Olivares et al.~(2013)]{Olivaresetal2013} Olivares J. S\'anchez L.~J., Ruelas-Mayorga A., Allen C., Costero R. \& Poveda A., 2013, AJ, 146, 106

\bibitem[O'Dell et al.~(2009)]{ODell2009}O'Dell, C.~R., Henney, W.~J., Abel, N.~P., et al. 2009, AJ, 137, 367

\bibitem[Pflamm-Altenburg \& Kroupa~(2006)]{Pflamm2006} {Pflamm-Altenburg}, J. \& {Kroupa}, P., 2006, MNRAS, 373, 295

\bibitem[Raghavan et al. (2010)]{Raghavanetal2010}Raghavan, D. McArthur, H.A., Henry, T.J. et al., 2010, ApJS, 190, 1

\bibitem[Sana et al. (2012)]{Sanaetal2012} Sana, H., de Mink, S.~E., de Koter, A., Langer, N., et al. 2012, Sci, 337, 444

\bibitem[Sana et al. (2014)]{Sanaetal2014} Sana, H., Le Bouquin, J.~B., Lacour, S., et al. 2014, ApJS, 215, 15

\bibitem[Underhill \& Hill~(1994)]{UnderhillHill2000} Underhill, A.~B. \& Hill, G.~M., 1994, ApJ, 432, 770


\end{thebibliography}
\end{document}